\documentclass[twocolumn,prl]{revtex4}

\usepackage{epsfig,amssymb,amsmath}

\begin{document}

\title{Transverse Oscillating Waves and the Effects of Capacitive Coupling in Long Intrinsic Josephson Junctions}

\author{Farzad Mahfouzi}

\address{Department of Physics and Astronomy, University of Delaware, Newark, DE 19716-2570, USA}

\begin{abstract}
In this paper we investigate the excitation of longitudinal and transverse plasma waves in intrinsic Josephson junctions. We consider the outermost branch of IV characteristic (IVc) in current biased case and try to find the conditions in which plasma waves can be excited. We change the parameters of the system and get the corresponding breakpoint current at which the plasma waves start to initiate. We present specifically the modes containing only transverse waves where we can have radiation. As a result we find the range of parameters that the system can radiate.

Pacs: 74.72.-h, 74.50.+r, 74.40.+k. Keywords: Plasma oscillation,
Terahertz Radiation, High Tc superconductors, Intrinsic
Josephson Junctions
\end{abstract}
\maketitle One of the topics in high Tc superconductors that has
recently attracted a great interests in the literature is the
observed terahertz wave emission from the layered structured
high-temperature superconductors.\cite{Science} It's widely
believed that cuprate  high-temperature superconductor like
Bi2Sr2CaCu2O8 (BSCCO) has a layered structure which leads to an
intrinsic Josephson junction (IJJ) effect. \cite{IJJ} Applying
current in $z$-direction we get AC Josephson oscillations that
works in terahertz range of of frequencies. Moreover, the
parameters in intrinsic Josephson junctions are controlled by the
atomic crystal structure rather than by leads and amorphous
dielectric layer in artificial JJs. This results in the existence
of different mechanisms of coupling among IJJs e.g. phonon
coupling, capacitive coupling,\cite{CCJJ} inductive coupling,
charge imbalance effect\cite{CIB}. In this paper we consider
inductive-capacitive coupling model where we can have
longitudinal and transverse oscillations. Radiation from this
system has already been investigated in literature
\cite{Radiation, one-j}. In this paper we investigate properties
of the transverse and longitudinal plasma waves excited by a
constant bias current in the multi-Josephson junctions. We
restrict ourselves to the two-dimensional IJJs, for simplicity,
and consider that the length of the system in y-direction is
small comparing to the magnetic penetration depth parallel to the ab-plane of the crystal. Therefore we assume the system to be homogeneous in y-direction.
The layers are stacked along the z-direction (c-axis). We
consider that the length of the junctions in x-direction $L_{x}$
is comparable to the magnetic penetration depth and the phase-differences of junctions along the
x-direction varies accordingly. It has been established that the
junctions in these multi-junction systems are coupled both
capacitively and inductively with each other. We show that the
plasma waves inside the junctions are excited parametrically by
the Josephson oscillations arising from the bias current.

The couplings can be incorporated into the dynamics of the
phase-differences in IJJs, using the generalized Josephson
relations. \cite{equ1, equ2, equ3}
$\frac{d\varphi_{i}}{dt}=\frac{2e}{\hbar}(V_{y,(i,i-1)}(x)-\alpha_C\nabla^{(2)}V_{y,(i,i-1)}(x))$
and $\frac{d\varphi_{i}}{dx}=\frac{2e\mu
d}{\hbar}(H_{y,(i,i-1)}(x)-\alpha_L\nabla^{(2)}H_{y,(i,i-1)}(x))$.
Where the parameters $\alpha_C$ and $\alpha_L$ are respectively
the capacitive and inductive coupling constants given by
$\alpha_C=\frac{\epsilon\mu^2}{sd}$ ,
$\alpha_L=\frac{\lambda^2_{ab}}{sd}$ with $s,d,\epsilon,\mu$
and$\lambda_{ab}$ being, respectively, the thickness of the
superconducting layer and insulating layer, the dielectric
constant of the insulating layers, the charge screening length of
the superconducting layers and the magnetic penetration depth
parallel to the ab-plane of the crystal. The equation of motion
for the gauge-invariant phase-differences $\varphi(x,t)$ in
dimension-less form is \cite{equ1, equ2, equ3}
\begin{equation}
\lambda^2(\widehat{L}^{-1})_{ij}\frac{\partial^2\varphi_j}{\partial
x^2}-(\widehat{C}^{-1})_{ij}\frac{\partial^2\varphi_j}{\partial
\tau^2}=\sin(\varphi_i)+\beta \dot{\varphi_i}-I/I_c
\end{equation}
where $\widehat{C}=1-\alpha_C\nabla^{(2)}$ and
$\widehat{L}=1-\alpha_L\nabla^{(2)}$,
$\lambda^2=\frac{\hbar}{2e\mu L_yI_c}$, $\tau=\omega_p t$ and
$\omega_p^2=\frac{\epsilon edI_c}{2\pi\hbar}$. The index
indicates the  $i$th junction and $\nabla^{(2)}$ is the
abbreviation of the second rank difference. This equation has
been obtained from conservation relation where the total
dimension-less current, $I/I_c$, can be written in terms of the
transverse current, $\frac{d}{L_y}\frac{\partial
H_{i,i+1}}{\partial x}$, displacement current,
$\frac{4\pi}{\epsilon d}\frac{\partial V_{i,i+1}}{\partial t}$,
Ohmic current, $\beta \dot{\varphi_i}$, and the Josephson
current, $\sin(\varphi_i)$.

In this model we consider that an electron in the position $x$ of
$i$th layer can only jump to the same position on the next layer
i.e. $J_{i,i+1}=\sin(\varphi_i)+\beta \dot{\varphi_i}$. We call
this as local hopping approximation which is an appropriate
approximation in cases of slow variation of phase difference in
$x$ direction. In this paper we use this approximation and
neglect the fast varying solutions.

\begin{figure}
 \centering
\includegraphics[height=60mm]{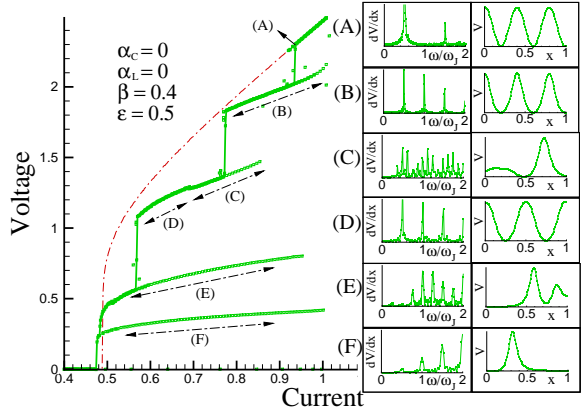}
\caption{(Color online) IV characteristic of one Josephson junction which is
equivalent to the case $\alpha_L=0, \alpha_C=0$. As the left inset of
the figure, we show the frequency spectrum of the derivative of
voltage with respect to $x$ and as the right inset of the figure we show
the distribution of voltage along the junction at an arbitrary time in the regions. The Current is defined in units of Josephson critical current and the Voltage is defined in units of $\hbar \omega_p/e$.} \label{IVC}
\end{figure}
In the following we are going to get some information about the
system of equations (1). It's well known that in the case of
short Josephson junctions the IVc has a multi branch structure
corresponding to different configurations of the rotating and
oscillating states among the junctions in $z$ direction. It has
been shown that neglecting the effects of boundary layers, the
number of branches is equal to the number of the Josephson
junctions.\cite{CCJJ} But what about the long Josephson junction
case? Does it lead to the same number of branches? In order to
answer this question, let us first solve the equation of motion
for one Josephson junction. The result of this solution is
presented in Fig. (\ref{IVC}). To obtain this result we have
considered $100$ points along the junction ($x$ direction). Then
using fourth order Rounge-Kuta algorithm we have solved the
system of equations in time. This result shows another kind of
branch structure which is well known and corresponds to different distributions of
phase along the junction. In general we can distinguish two types
of solutions corresponding to flux-flow mode and oscillating
distributions. It's clear from this result that if we start from outermost branch we reach first the oscillating mode and then to the flux-flow mode. To clarify the situation we specify some regions of the IVc
by letters $(A)$, $(B)$, $(C)$, $(D)$, $(E)$, $(F)$ and present,
as left inset, the FFT analysis of the time dependence of
$\frac{V(x=0.02)-V(0.01)}{0.01}$ in units of Josephson frequency
and as right inset, the distribution of voltage along the
junction. We can see three, two and one flux-flow mode states at
the regions $(C)$, $(D)$ and $(F)$ respectively and standing
oscillating waves at the regions $(B)$ and $(D)$. Although the
number of fluxons can be seen directly from the distribution of
voltage, we can also find it from the number of peaks in unit
range of the FFT results i.e. $N_{peaks}=2^{N_{fluxon-antifluxons}}$.

Let us now consider the multi layered structure and find the
number of branches in IVc in cases of Long Josephson junctions.
From above discussion we find that there are at least two
mechanisms of branching in IVc $i.e. $(1) different configurations
of the rotating and oscillating states among the junctions in $z$
direction and (2) different distributions of phase along the
junction in $x$ direction. Should we consider both effects the
number branches exceeds the number of junctions which is
apparently not in accordance with experimental results. In the
following we see that in the case of high numbers of junctions
the number of branches tends to the number of junctions. We show
this by noting that at large number of junctions, upon creation of longitudinal plasma wave at the breakpoint current the system immediately transits to another branch containing less resistive junctions. Therefore, should the system starts from an oscillating mode with
longitudinal waves, the system does not reach other resonance
regions corresponding to some other transverse modes and so the
additional branches due to inductive coupling in $x$-direction does not appear. But what if the system in
the outermost branch starts from a state with only transverse mode
(without longitudinal mode)$i.e.$ (n,0)? It's clear that, in this case, upon decreasing the current, the system continues to be in this state until it reaches to another resonance region which contains longitudinal plasma wave and then the system jumps to another branch corresponding to less number of resistive states.

From Fig. (\ref{IVC}) we learned that the flux-flow modes appear far from the breakpoint current. Additionally we know that in multi Josephson junction case, upon creation of longitudinal plasma waves the system jumps to another state with less number of resistive junctions. In the following, we neglect the flux-flow modes and consider only the region close to breakpoint current of outermost branch in IVc. Later in this paper we can see that, if we consider the longitudinal coupling, the system in fact hardly reaches to the flux-flow regions and therefore the flux-flow modes play no effective role in multi Josephson junctions.
As a result we start from points $(A)$ and $(B)$ in Fig. (\ref{IVC}) and investigate their characteristics.
Point $(A)$ represents breakpoint
current and the system at this current enters into the parametric resonance region. Throughout this paper we consider this point only and discuss about its features. In Fig. (\ref{IVC}) we see that, the region $(B)$ which we call radiation branch, has close characteristics to the point $(A)$ and also it appears after the breakpoint current. In the following we discuss that this branch might be observable in IVc.

\begin{figure}
 \centering
\includegraphics[height=70mm]{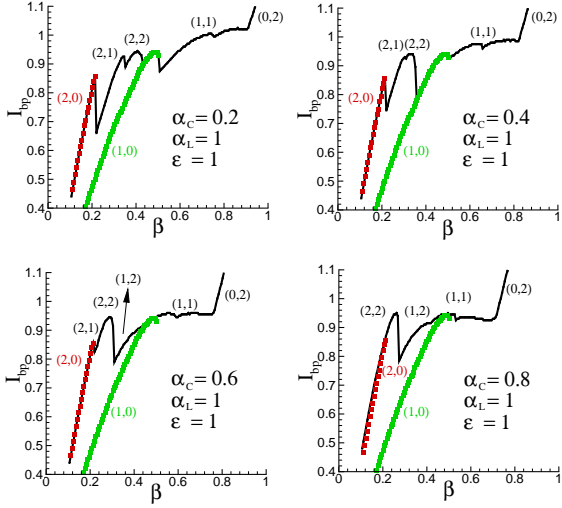}
\caption{ (Color online)$\beta$ dependence of the breakpoint current in units of critical current at
$\alpha_L=1, \varepsilon =1$ and different capacitive coupling
($\alpha_C$)}. \label{I-c-dep}
\end{figure}
By the fact that on the outermost branch all of the junctions are
in the resistive (rotating) state, we neglect the effect of
boundaries (large number of junctions) and consider that initially,
phases of the junctions are the same. We consider also that this
phase does not depend on the position $x$ i.e.
$\varphi_i(x,t)=\phi(t)$. Then we test the stability of such a
state by adding a small term to the phase i.e.
$\varphi_i(x,t)=\phi(t)+\delta_i(x,t)$. Linearizing the equation
with respect to this term and then making Fourier transformation
with respect to $x$ and $i$ leads
\begin{equation}
\frac{1}{C(k_z)}\frac{d^2\delta(k_x,k_z)}{d\tau^2}+\beta
\frac{d\delta(k_x,k_z)}{d\tau}+(\varepsilon\frac{n^2}{L(k_z)}+\cos(\phi(\tau)))\delta=0
\end{equation}
\begin{equation}
\frac{d^2\phi}{d\tau^2}+\beta \frac{d\phi}{d\tau}+\sin(\phi)=I/I_c
\end{equation}
where $C(k_z)=1+2\alpha_C(1-\cos(k_z))$,
$L(k_z)=1+2\alpha_L(1-\cos(k_z))$, $\varepsilon=\pi
\lambda^2/L_x^2$ and $n$ is an integer number. Due to absence of
external magnetic field the derivative of phase at the boundaries
of the junctions are zero. This boundary condition leads to
$k_x=n\pi$. On the other hand, considering a periodic boundary
condition for $z$-direction we get $k_z=2m\pi/N$. Where $m$ is
also an integer number and N is the number of junctions. $n$ and
$m$ represent the transverse and longitudinal wave numbers
respectively.
\begin{figure}
 \centering
\includegraphics[height=70mm]{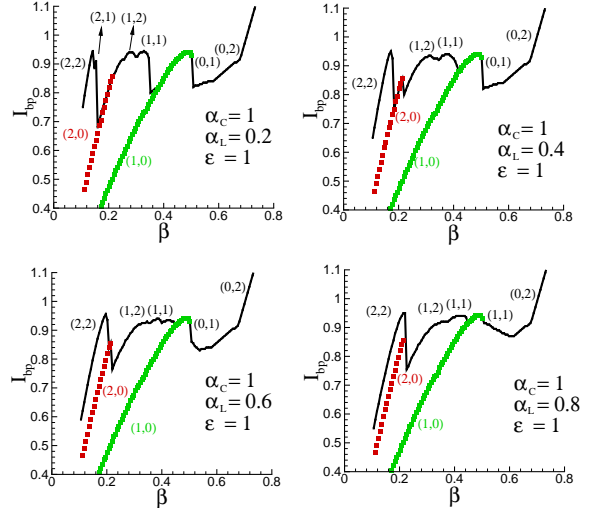}
\caption{ (Color online)$\beta$ dependence of the breakpoint current in units of critical current at
$\alpha_C=1, \varepsilon =1$ and different inductive coupling
($\alpha_L$)}. \label{I-L-dep}
\end{figure}

\begin{figure}
 \centering
\includegraphics[height=70mm]{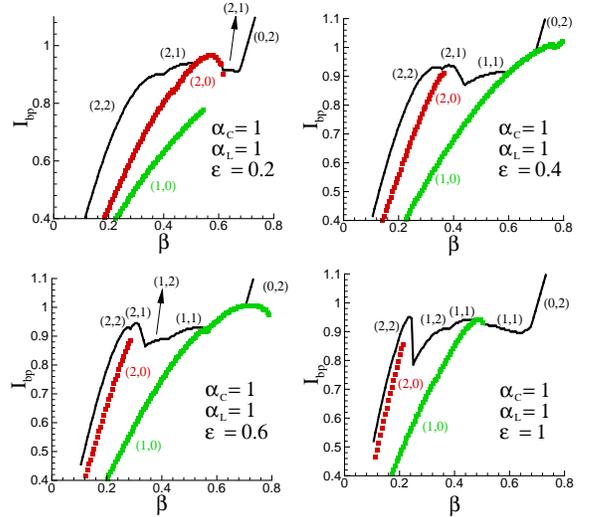}
\caption{ (Color online)$\beta$ dependence of the breakpoint current in units of critical current at
$\alpha_C=1, \alpha_L =1$ and different $\varepsilon$}.
\label{I-g-dep}
\end{figure}

We find that the this initial condition is stable at the
big currents and unstable bellow some critical current which we
call breakpoint current. We find the breakpoint current at
different parameters of the system. Fig. \ref{I-c-dep} shows
breakpoint current versus $\beta$ at different capacitive
coupling parameters. The black curve shows the maximum of
breakpoint current for different modes $(n,m)$ where $n$ and $m$
are taking the values $0,1,2$. Where $m_{max}=2$ means that the
number of Josephson junctions is 2 and $n_{max}=2$
means that we are neglecting fast varying solutions. Additionally one can see that bigger values of $n$
appear at smaller $\beta$. Therefore, to see the results
corresponding to bigger $n$ we need to choose smaller $\beta$ which is out of scope of this paper, because, we are neglecting fast varying solutions. 
In Figs. \ref{I-c-dep},
\ref{I-L-dep}, \ref{I-g-dep} we show especially the modes that
longitudinal waves are absent i.e. the oscillation that can lead
to radiation. We note that the parts of these curves that
coincide with the black curve are observable only. We note also
that increasing the number of junctions smears the
$\beta$-dependence of the breakpoint current.

\begin{figure}
 \centering
\includegraphics[height=70mm]{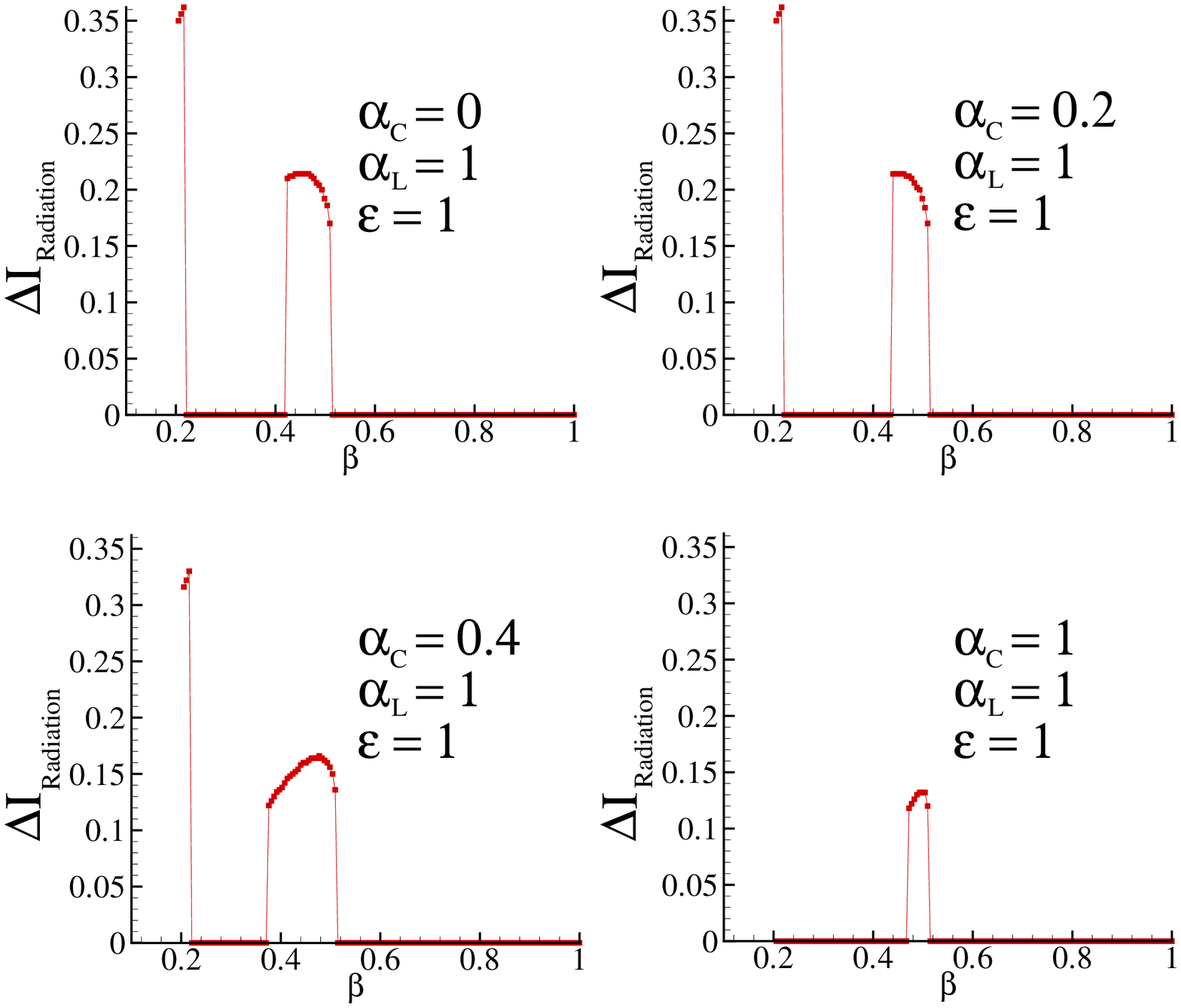}
\caption{(Color online)Width of radiation range on the outermost branch of IVc
($\Delta I_{Radiation}$) versus $\beta$ at different capacitive
coupling $\alpha_C$} \label{wfrequency}
\end{figure}

From experimental point of view, it might be interesting to know
the width of the region in IVc that there's electromagnetic
radiation and it's dependence on the parameters of the system.
The resonance width is defined by the difference between
breakpoint current and jump point current at which the system
goes to another branch. We have already shown that in the
presence of longitudinal plasma waves the width of the resonance
region is getting zero with increasing the number of
junctions.\cite{bp-ccjj} In case of transverse plasma waves the
situation is different. In addition to the fact that the breakpoint
current in this case doesn't depend on the number of junctions, result of simulation for one
Josephson junction shows also that the return current doesn't depend
strongly on the mode of transverse plasma waves. This leads to
the conclusion that jump point current can be obtained from the
breakpoint of the modes with the longitudinal plasma waves.
Therefore we can get the range of current that we have only
transverse modes and and can be touched from the outermost
branch. We can see the result of this calculation in Fig. \ref{wfrequency}.
This model predicts that if we get the branch
corresponding to radiation (only transverse waves) in the return
current part of the outermost branch, we can then increase the
current and have radiation until the end of the branch. But one should note that the end of branch in this case is not critical current ($I=I_c$). In addition, if we include the dissipation due to the radiation power, it will lead to increase in voltage in IVc and therefore decrease in the length of the radiation branch. We note
that although increasing the number of junctions has not strong
effect on the peak of the $\beta$-dependence of $\Delta
I_{Radiation}$ but it decreases the range of $\beta$ that $\Delta
I_{Radiation}$ is nonzero. This shows that increasing the number
of junctions makes the situation more difficult to get the
radiation region.

Now that we know the condition for electromagnetic radiation one
question arises; What is the frequency of oscillation at the
breakpoint current and breakpoint region ($\omega_{Jbp}$)? Results
of our numerical calculations $e.g. $ $(A)$ point in Fig.
\ref{IVC}, shows that the Josephson frequency at the breakpoint
in IVc follows the $(M+\frac{1}{2})\omega_{Jbp}$ relation, where
$M$ is integer number. We note that this result corresponds to
the breakpoint of outermost branch which is not stationary. The
stationary solution which can be obtained from the original
nonlinear equation and we call radiation branch, contains frequencies equal to
$\frac{1}{2}M\omega_{Jbp}$.
\begin{figure}
 \centering
\includegraphics[height=70mm]{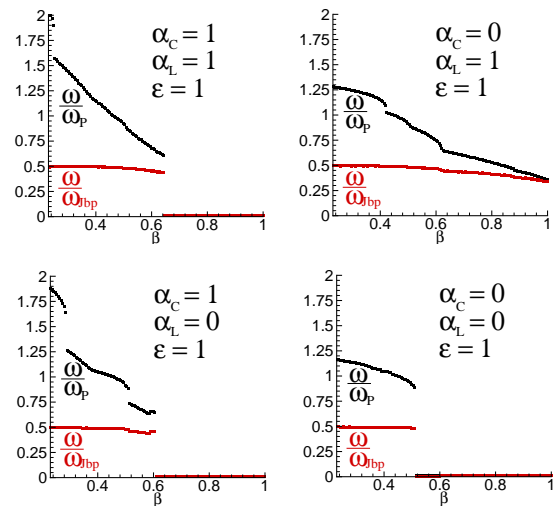}
\caption{(Color online)Minimum frequency of the plasma wave spectrum in units of Josephson plasma frequency ( black
curves) and breakpoint Josephson frequency (red curves).} \label{frequency}
\end{figure}
In Fig. \ref{frequency} we show the first frequency of the plasma
waves ($M=0$) in units of Josephson plasma frequency ( black
curves) and in units of breakpoint Josephson frequency (red
curves). We can see that the frequency of plasma waves is close
to half of the Josephson frequency at the breakpoint. The
deviation is due to the fact that the system is current biased.

We conclude that upon decreasing the bias from the fully
resistive state, the system enters into resonance between the
Josephson frequency and the plasma waves. As a result we have
stationary plasma waves that can contain longitudinal and
transverse waves. Should the longitudinal wave exist, some of the
junctions immediately switch back from the resistive into
superconducting states. In case of transverse waves without
longitudinal one, an additional branch, that we call the
radiation branch, appears and the system starts to radiate
electromagnetic power. Upon decreasing current on the radiation
branch the system finally enters into the resonance regions
corresponding to other modes of plasma waves that yields a
transition to another state with less resistive states. Therefore
in the long intrinsic Josephson junction systems where both
longitudinal and transverse waves exist, the branching in IVc is
mainly due to longitudinal waves and the only observable effect of
transverse waves is accompanied by radiation.

The author would like to thank Yu. Shukrinov, B. K. Nikolic and M.
Machida for their helpfull discussions.

\end{document}